\documentclass[
aps,%
12pt,%
final,%
notitlepage,%
oneside,%
onecolumn,%
nobibnotes,%
nofootinbib,% In the current version of REVTeX, when this option is on,
superscriptaddress,%
noshowpacs,%
centertags]%
{revtex4}

\usepackage{graphicx}
\usepackage{color}

\newcommand{\pT}{\ensuremath{p_{\rm t}}} 
 
\newcommand{\sqsNN}{\sqrt{s_{_{NN}}}}
\begin{document}
\selectlanguage{english} % For a paper in English

\title{Recent results from ALICE}
\author{\firstname{Yuri} \surname{Kharlov, for the ALICE collaboration}}
\affiliation{Institute for High Energy Physics, Protvino, 142281 Russia}

\begin{abstract}
  The ALICE experiment at the LHC has collected wealthy data
  in proton-proton and lead-lead collisions. An overview of recent
  ALICE results is given in this paper. Hadron spectra
  measured in pp collisions at $\sqrt{s}=0.9$, 2.76 and 7~TeV are
  discussed. Properties of hot nuclear matter produced in Pb-Pb collisions
  at $\sqrt{s_{_{NN}}}=2.76$~TeV, revealed via many observables
  measured with the ALICE experiment, are shown.
\end{abstract}

\maketitle
%-----------------------------------------------------------------------------
\section{Introduction}

The ALICE experiment was designed to study interactions of heavy ions
at the LHC. This goal determines the unique performance of the ALICE
detectors to reconstruct events with very high multiplicity and to
measure spectra of identified hadrons, electrons, photons, muons in a
wide energy range. 

The data collected with the ALICE experiment with pp collisions at
$\sqrt{s}=7$~TeV in 2010--2011, consist of a minimum-bias sample with
integrated luminosity $\int{\cal L}dT = 16~\mbox{nb}^{-1}$ and a
sample recorded with rare-event triggers with integrated luminosity
$\int{\cal L}dT = 4.9~\mbox{pb}^{-1}$. Rare-event triggers implemented
pp collisions were based in EMCAL, PHOS and MUON detectors.  Limited
data samples with the proton beams at collision energies
$\sqrt{s}=0.9$ and 2.76~TeV have been also recorded with integrated
luminosities $\int{\cal L}dT = 0.14\mbox{~and~}1.3~\mbox{nb}^{-1}$
respectively.

The LHC has delivered heavy-ion collisions at the center-of-mass
energy $\sqsNN=2.76$~TeV to the ALICE experiment in 2010 with
integrated luminosity $\int{\cal L}dT = 10~\mu\mbox{b}^{-1}$, and the
data set of 2011 exceeded the previous one by an order of
magnitude. Data taking of heavy ion collisions recorded in 2010 was
dominated by the minimum bias trigger. In 2011, a fraction of minimum
bias events was suppressed in favor of the triggers on the most
central and semi-central events, as well as rare events which selected
events with high-energy clusters in the electromagnetic calorimeters,
muon tracks in the muon spectrometer, ultra-peripheral
collisions. 

%-----------------------------------------------------------------------------
\section{QCD tests in proton-proton collisions}

Properties of hot nuclear matter produced in heavy ion collisions are
studied via a comprehensive set of observables. As a reference,
similar observables are measured in proton-proton collisions. ALICE is
performing detailed studies of hadron production spectra in pp
collisions at all center-mass energies provided by the LHC. Apart of
being a reference for heavy ion collisions, pp collisions is
considered as a powerful tool for QCD studying. Advance particle
identification capabilities \cite{Aamodt:2008zz} and a moderate
magnetic field ($B=0.5$~T) allow to measure a variety of hadron
spectra in a wide momentum range.

Identified charged hadron production in mid-rapidity are measured by
the central tracking system consisting of the Inner Tracking System
detector (ITS), Time Projection Chamber (TPC), Time-of-Flight detector
(TOF) and a Cherenkov High-Momentum Particle Identification detector
(HMPID). Each of these detectors provide particle identification in
different complimentary momentum ranges, which allows to measure the
spectra in a wide $\pT$ range.  ALICE has already published production
spectra of $\pi^\pm$, $K^\pm$, p, $\bar{\mbox{p}}$ in pp collisions at
$\sqrt{s}=0.9$~TeV \cite{PIDhadron900GeV}, and has reported
preliminary results on those spectra in pp collisions at
$\sqrt{s}=7$~TeV \cite{PIDhadron7TeV}. Similar to charged hadrons,
ALICE is able to measure neutral meson spectra by complimentary and
redundant methods which ensures the result validity.  Neutral pion and
$\eta$-meson spectra were measured in pp collisions at all three LHC
energies by the Photon Spectrometer (PHOS) which detected real photons
and by the central tracking system which identifies photons converted
to $e^+e^-$ pairs on the medium of the inner ALICE detectors
\cite{pp-pi0}. Combined analysis of all ALICE detectors allowed to
measure spectra of resonance production, in particular strange and
charmed hadrons. 

Results obtained by the ALICE on hadron production in pp collisions
show a gradual increase of the mean transverse momentum with collision
energy. Observed ratio between antiprotons and protons suggests that
baryon-antibaryon asymmetry is restoring at high energies. Comparison
of all measured spectra with Monte Carlo event generators and with
next-to-leading perturbative QCD calculations demonstrates that no
model can correctly describe spectra of hadron production at LHC
energies.

%.............................................................................
\section{Global event properties in Pb-Pb collisions}

\begin{figure}[ht]
  \centering
  \includegraphics[width=0.60\hsize]{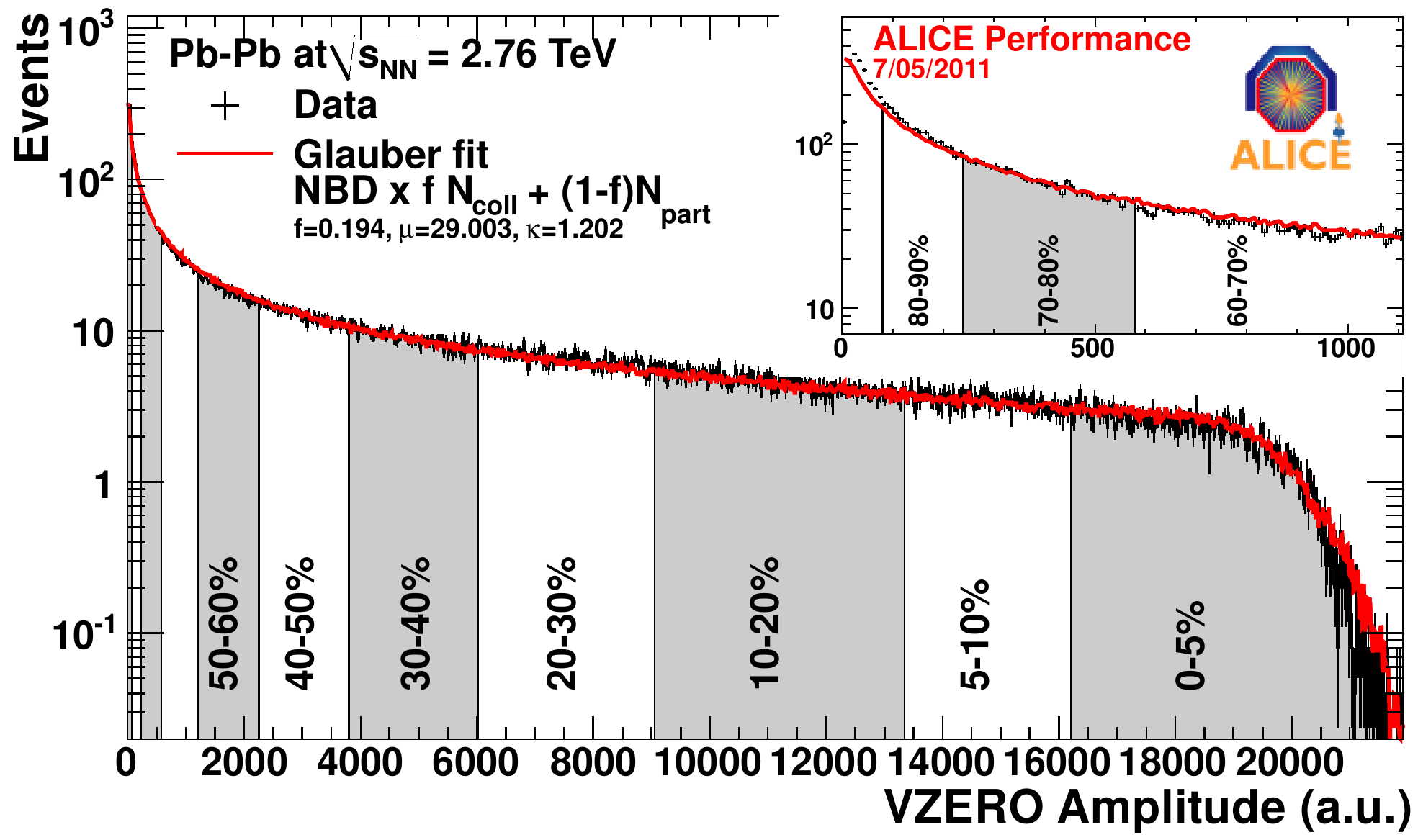}
  \caption{Centrality determination in ALICE. Glauber model fit to the
    VZERO amplitude with the inset of a zoom of the most peripheral
    region.}
  \label{fig:PbPb-centrality}
\end{figure}

Centrality of the collision, directly related to the impact parameter
and to the number of nucleons $N_{\rm part}$ participating in the
collision, allows to study particle production versus density of
the colliding system.  Various ALICE detectors measure collision
centrality, among them the best accuracy is achieved with the
scintillator hodoscope VZERO covering pseudorapidity ranges $2.8 <
\eta < 5.1$ and $-3.7 < \eta < -1.7$. Distribution of the sum of
amplitudes in VZERO in minimum bias Pb-Pb collisions is shown in
Fig.\ref{fig:PbPb-centrality} \cite{bib:PbPb-dNdy}. Centrality
classes were defined by the Glauber model, and the fit of the Glauber
model to the data is shown by a solid line in this plot.

\begin{figure}[ht]
  \centering
  \includegraphics[width=0.60\hsize]{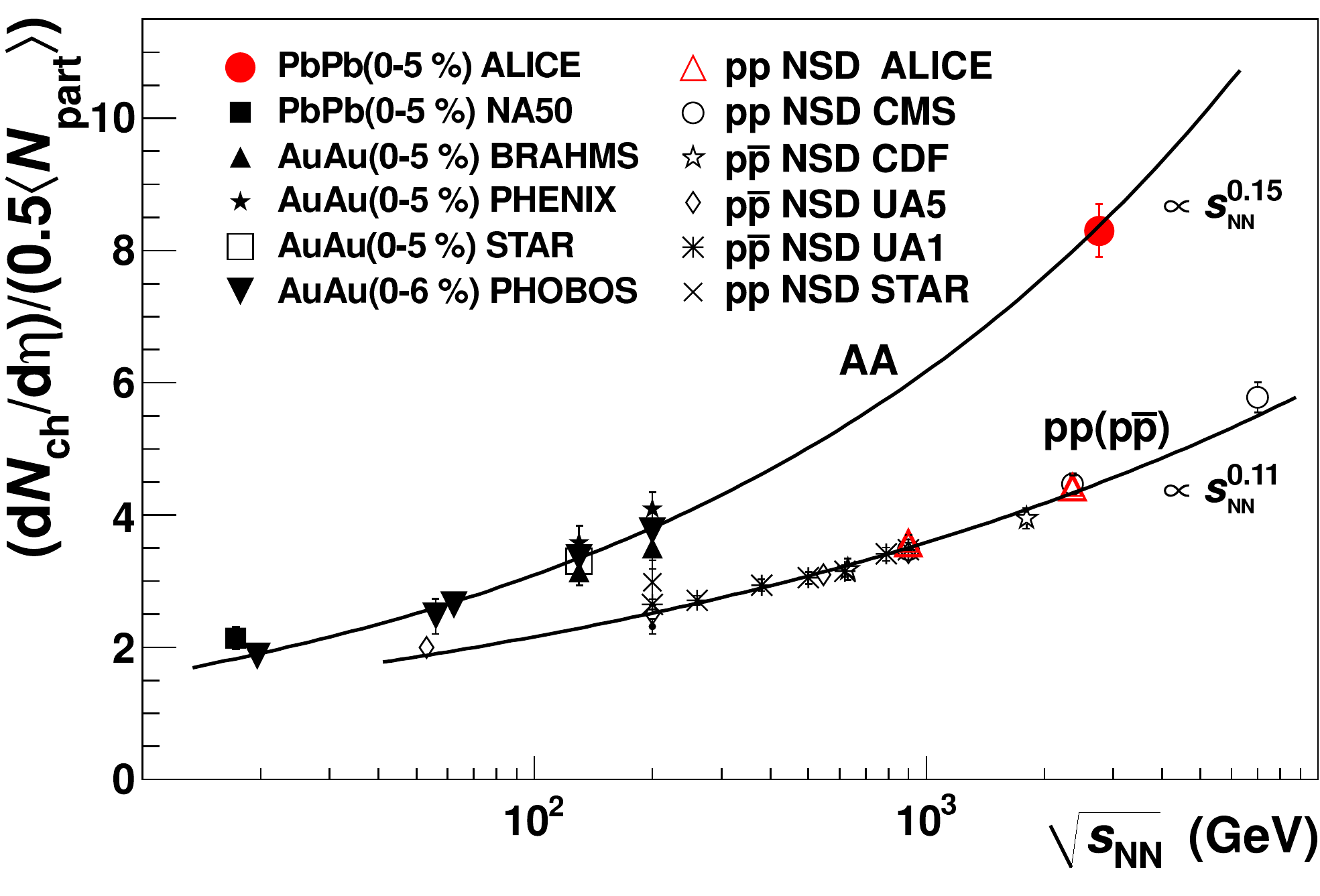}
  \hfill
  \caption{Charged track density $dN/d\eta$ in pp and AA collisions vs
    collision energy.}
  \label{fig:PbPb-dNdeta}
\end{figure}

Charged particle multiplicity in Pb-Pb collisions at $\sqsNN=2.76$~TeV
and its dependence on the collision centrality was measured with the
Silicon Pixel Detector (SPD), two innermost layers of the barrel
tracking system covering the pseudorapidity range $|\eta|<1.4$. The
charged particle density, normalized to the average number of
participants in a given centrality class, $dN_{\rm
  ch}/d\eta/\left(\langle N_{\rm part} \rangle \right)$, measured by
ALICE, was compared with similar measurements at lower energies at
RHIC and SPS (Fig.\ref{fig:PbPb-dNdeta}) \cite{bib:PbPb-dNdy_central}.
In the most central events (centrality $0-5\%$) at LHC energy the
charged particle density was found to be $dN_{\rm ch}/d\eta=1601\pm
60$ which is, being normalized to the number of
participants, is 2.1 times larger than the charged particle density
measured at RHIC at $\sqsNN=200$~GeV and 1.9 times larger than that in
pp collisions at $\sqrt{s}=2.36$~TeV. 

%.............................................................................
\section{Collective phenomena in heavy ion collisions}

The initial anisotropy of nuclei non-central collisions leads to
anisotropic distribution on initial matter in the overlapping
reagion. During evolution of the matter, the spatial asymmetry of
initial state is converted to an anisotropic momentum
distribution. The azimuthal distribution of the particle yield can be
described by a Fourier series of a function of the angle between the
particle direction $\varphi$ and the reaction place $\Psi_{\rm RP}$.
The second coefficient of this series, $v_2$, is referred to as
elliptic flow. Theoretical models, based on relativistic hydrodynamics
\cite{bib:hydro-v2_Kestin,bib:hydro-v2_Niemi}, successfully described
the elliptic flow observed at RHIC and predict its increase at LHC
energies from 10\% to 30\%.

\begin{figure}[ht]
  \centering
  \includegraphics[width=0.60\hsize]{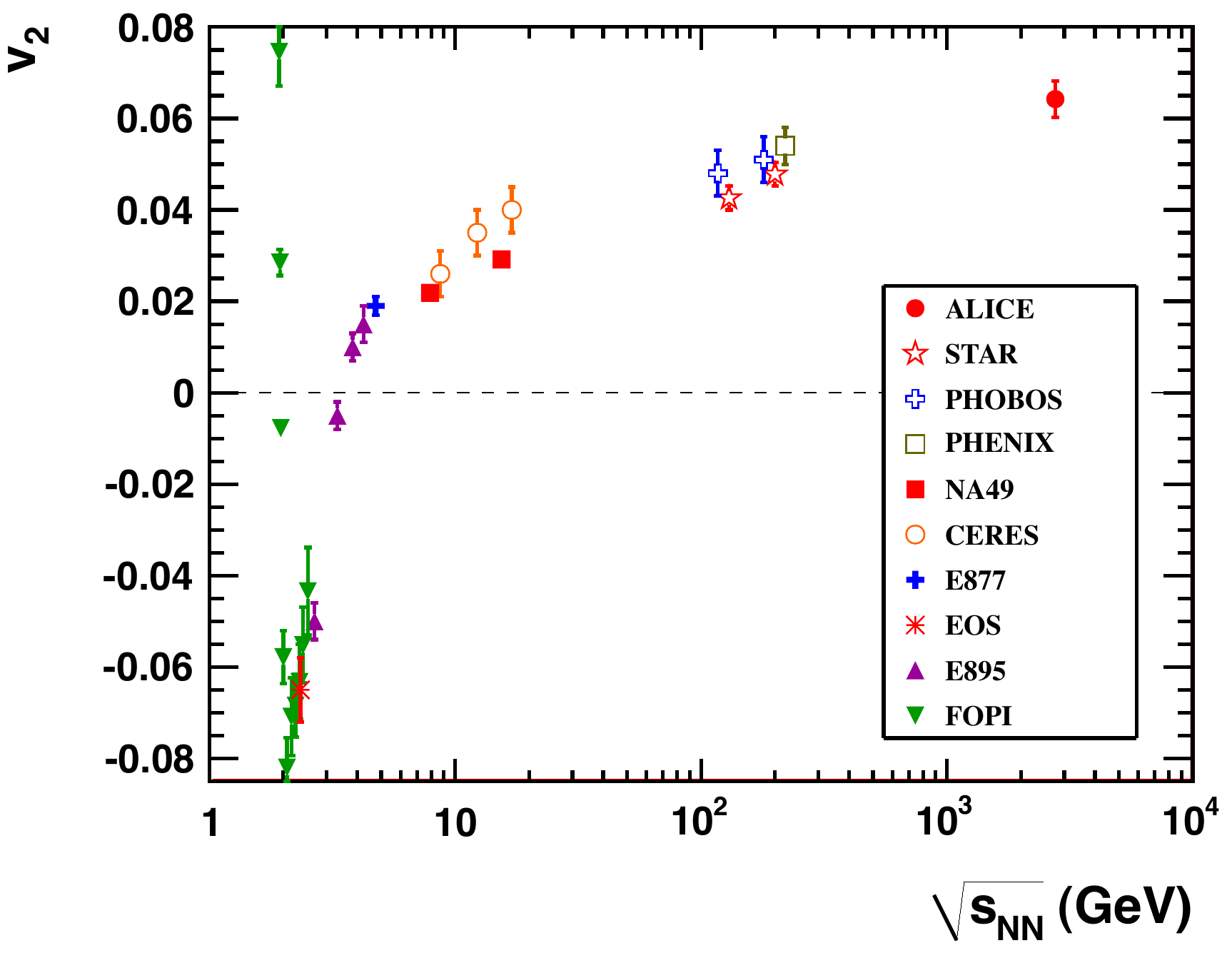}
  \caption{Azimuthal flow $v_2$ of charged particles measured by the
    ALICE in Pb-Pb collisions at $\sqsNN=2.76$~TeV in comparison with
    the lower-energy experiment results.}
  \label{fig:PbPb-v2}
\end{figure}

The first measurements of elliptic flow of charged particles in Pb-Pb
collisions at $\sqsNN=2.76$~TeV were reported by ALICE in
\cite{bib:ALICE-v2}. Charged tracks were detected and reconstructed in
the central barrel tracking system, consisting of ITS and
TPC. Comparison of elliptic flow integrated over $\pT$ measured by the
ALICE and lower-energy experiments is shown in
Fig.\ref{fig:PbPb-v2}.
The observed trend of $v_2$ vs $\sqsNN$ confirms model
expectations that the value of $v_2$ in Pb-Pb collisions at
$\sqsNN=2.76$~TeV increases by about 30\% with respect to $v_2$ in
Au-Au collisions at $\sqsNN=0.2$~TeV.

The results of global event properties and collective expantion
studied by ALICE, studied via the azimuthal anisotropy and intensity
interferometry of identical particles \cite{bib:HBT}, indicate that
the fireball formed in nuclear collisions at the LHC is hotter, lives
longer, and expands to a larger size at freeze-out as compared to
lower energies.

%.............................................................................
\section{Strangeness production in heavy ion collisions}

\begin{figure}[ht]
  \includegraphics[width=0.48\hsize]{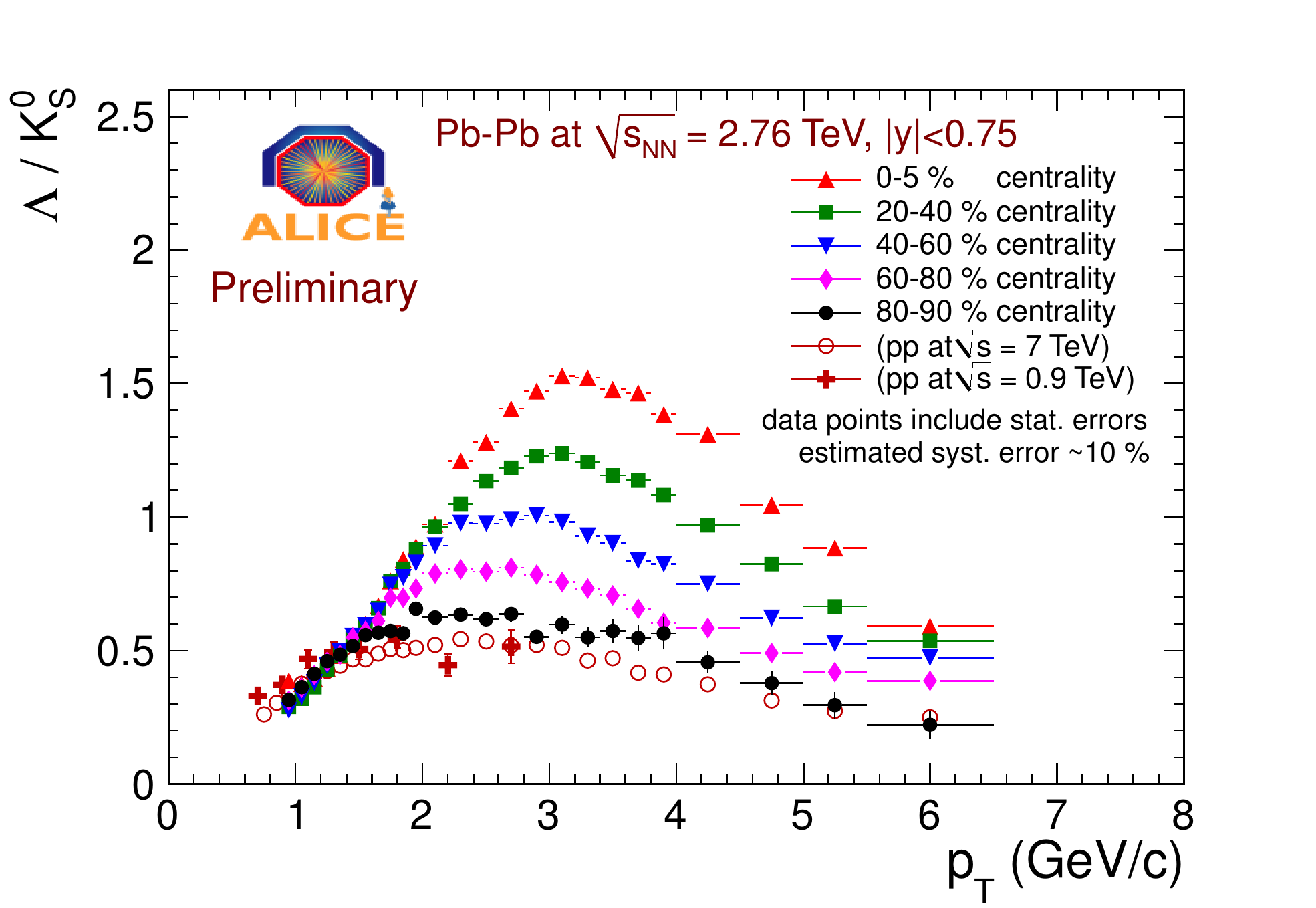}
  \hfill
  \includegraphics[width=0.48\hsize]{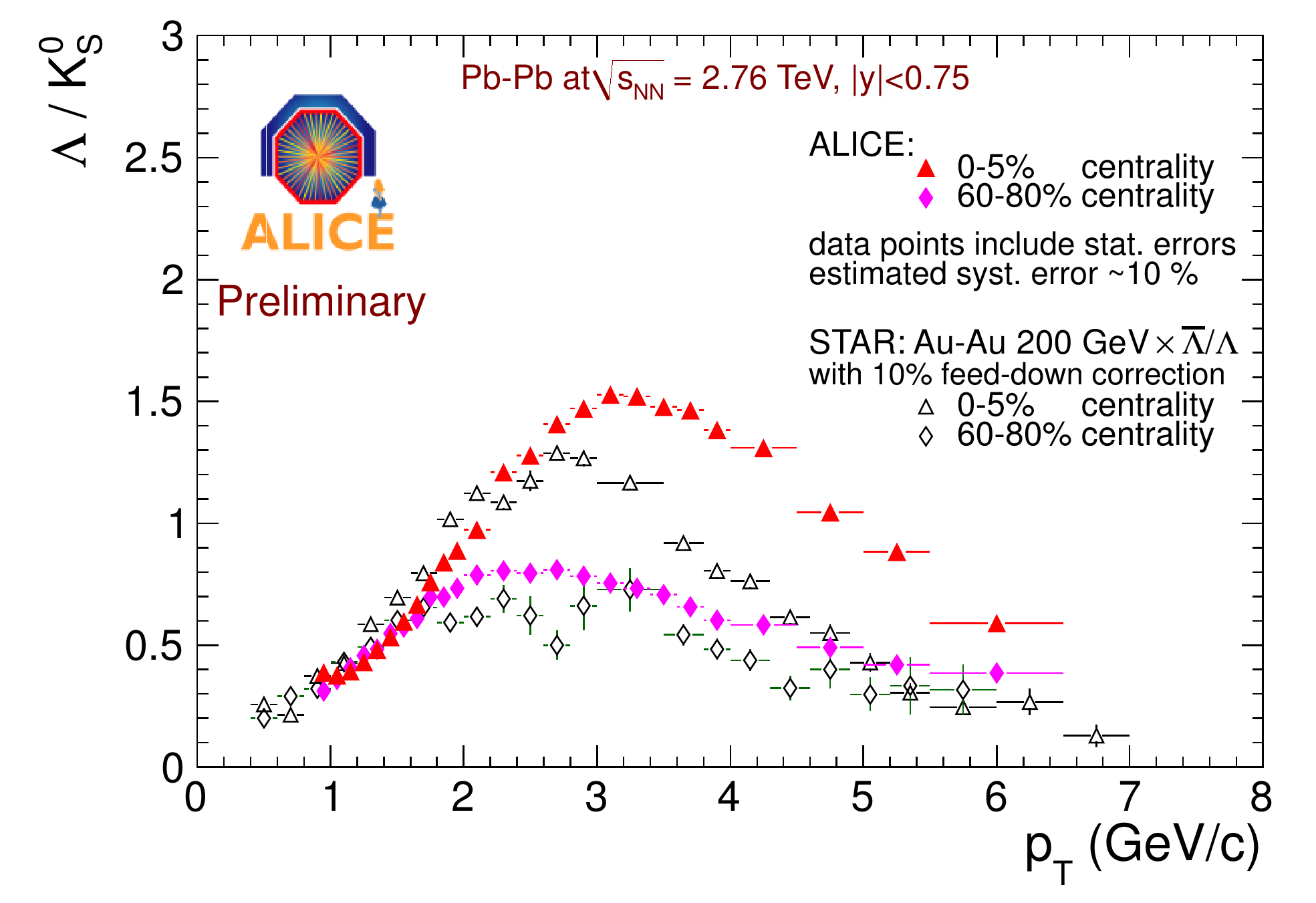}
  \caption{Ratio $\Lambda/K^0_S$ in Pb-Pb collisions at
    $\sqsNN=2.76$~TeV in different centralities (left) and
    comparison of this ratio at LHC and RHIC in centralities $0-5\%$
    and $60-80\%$ (right).}
  \label{fig:PbPb-Lambda_K0s}
\end{figure}

Strange particle production has been considered as a probe of strongly
interacting matter by heavy-ion experiments at AGS, SPS and RHIC. We
have already demonstrated that ALICE, due to its powerful particle
identification technique, has measured strange particle spectra in pp
collisions. Similar analysis was performed on the Pb-Pb data collected
in 2010. Comparison of strange meson and baryon production is
illustrated by the $\Lambda/K^0_S$ ratio measured by ALICE in
different centrality classes (Fig.\ref{fig:PbPb-Lambda_K0s},
left). This ratio in peripheral Pb-Pb collision is similar to that one
measured in pp collisions, but it grows with centrality,
increasing the value of 1.5 in the most central collisions. The
qualitative behaviour of this ratio on $\pT$ at the LHC collision
energy is similar to the ratio measured at RHIC by the STAR experiment
(Fig.\ref{fig:PbPb-Lambda_K0s}, right). An enhancement of strange and
multi-strange baryons ($\Omega^-$, $\bar{\Omega}^+$,
$\Sigma^-$,$\bar{\Sigma}^+$ ) was obsevred in heavy-ion collisions by
experiments at lower energies, and was confirmed by ALICE at LHC
energy \cite{ALICE-Hippolyte}. It was also shown that multi-strange
baryon enhancement scales with the number of participants $N_{\rm
  part}$ and decreases with the collision energy.

The large yield of strange, and especially multi-strange baryons in
heavy-ion collision was observed earlier at SPS and RHIC. This effect
supports predictions of quark-gluon plasma formation which assumed
that strange antiquarks are as abundant as light antiquarks in quark
matter. The strange quark phase space becomes fully equilibrated, and
therefore all strange hadrons are produced more abundantly. An
overview of strangeness production in heavy ion collisions can be
found in \cite{bib:Muller2011}.

%.............................................................................
\section{Parton energy loss in medium}

Final-state partons produced at the initial stage of nucleus-nucleus
collision, pass through medium with multiple secondary
interactions. Energy loss by partons depends on density and
temperature of the QCD medium. Hadrons produced in fragmentation of
these partons should be suppressed compared to expectations from an
independent superposition of nucleon-nucleon collisions.  The strength
of suppression of a hadron $h$ is expressed by the nuclear
modification factor $R_{AA}$, defined as a ratio of the particle
spectrum in heavy-ion collision to that in pp, scaled by the number of
binary nucleon-nucleon collisions $N_{\rm coll}$:
\begin{equation}
R_{AA}(\pT) = \frac{(1/N_{AA})d^2N_h^{AA}/d\pT d\eta}
                  {N_{\rm coll}(1/N_{pp})d^2N_h^{pp}/d\pT d\eta}.
\end{equation}
Experiments at RHIC reported that hadron production at high transverse
momentum in central Au-Au collisions at a center-of-mass energy per
nucleon pair $\sqsNN=200$~GeV is suppressed by a factor $4-5$ with
respect to pp collisions.  At the larger LHC energy, the density of
the medium is expected to be higher than at RHIC, leading to a larger
energy loss of high-$\pT$ partons. However, the hadron production
spectra are less steeply falling with $\pT$ at LHC than at RHIC which
would reduce the value of $R_{AA}$ for a given value of the parton
energy loss.

\begin{figure}[ht]
  \centering
  \includegraphics[width=0.50\hsize]{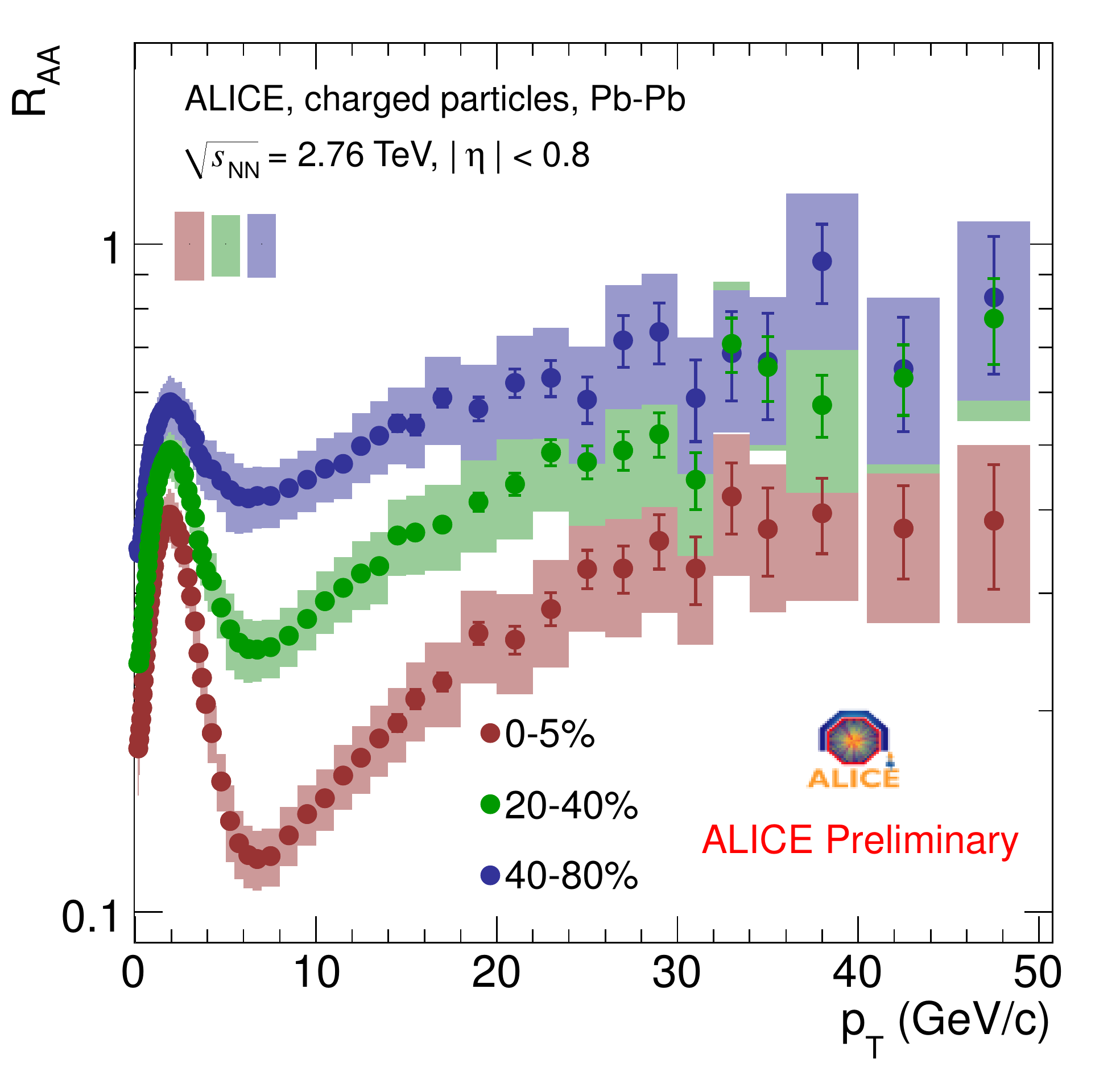}
  \caption{Nuclear modification factor $R_{AA}$ of charged particles.}
  \label{fig:PbPb-RAA_charged}
\end{figure}

ALICE has measured the nuclear modification factor $R_{AA}$ for many
particles. All charged particles, detected in the ALICE central
tracking system (ITS and TPC), show a spectrum suppression
\cite{Otwinowski:2011gq} which is qualitatively similar to that
observed at RHIC (Fig.\ref{fig:PbPb-RAA_charged}).
However, quantitative comparison with RHIC demonstrates that the
suppression at LHC energy is stronger which can be interpreted by a
denser medium. Benefiting from particle identification which has been
already mention earlier in this paper, ALICE has measured suppression
of various identified hadrons, which provides experimental data for
studying the flavor and mass dependence of the spectra suppression.

\begin{figure}[ht]
  \includegraphics[width=0.58\hsize]{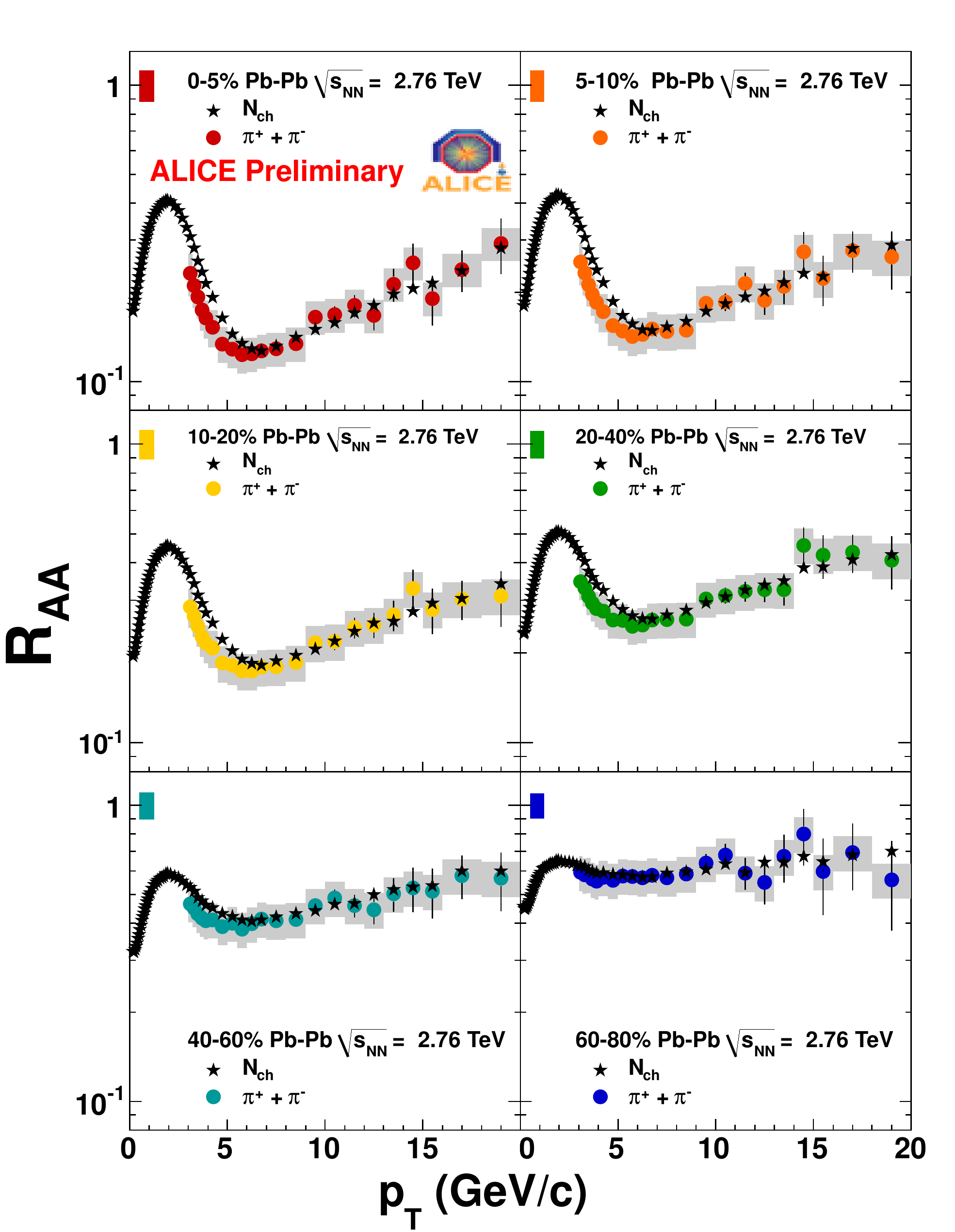}
  \caption{Nuclear modification factor $R_{AA}$ of charged pions.}
  \label{fig:PbPb-RAA_pions}
\end{figure}

\begin{figure}[ht]
  \includegraphics[width=0.48\hsize]{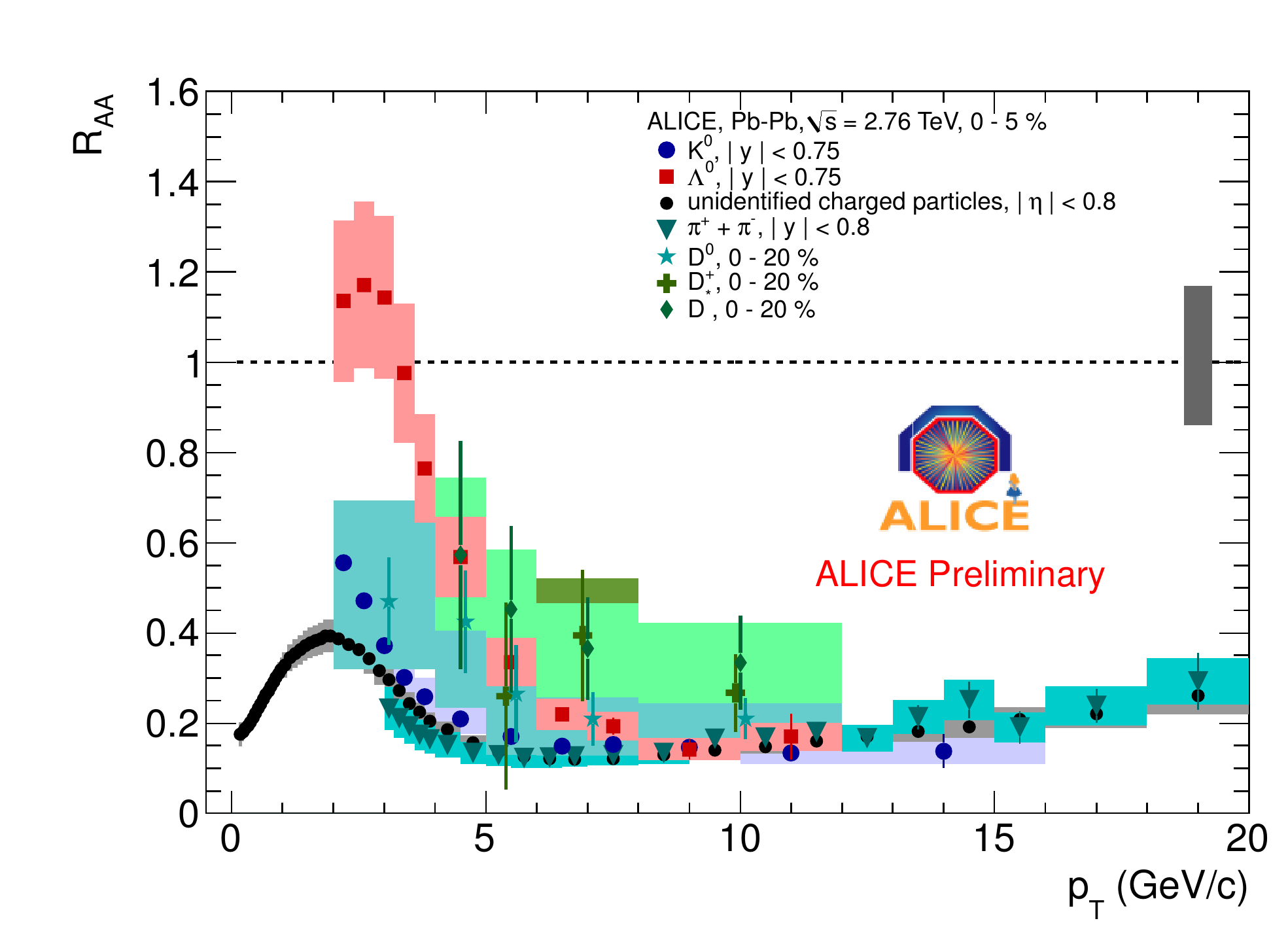}
  \hfill
  \includegraphics[width=0.48\hsize]{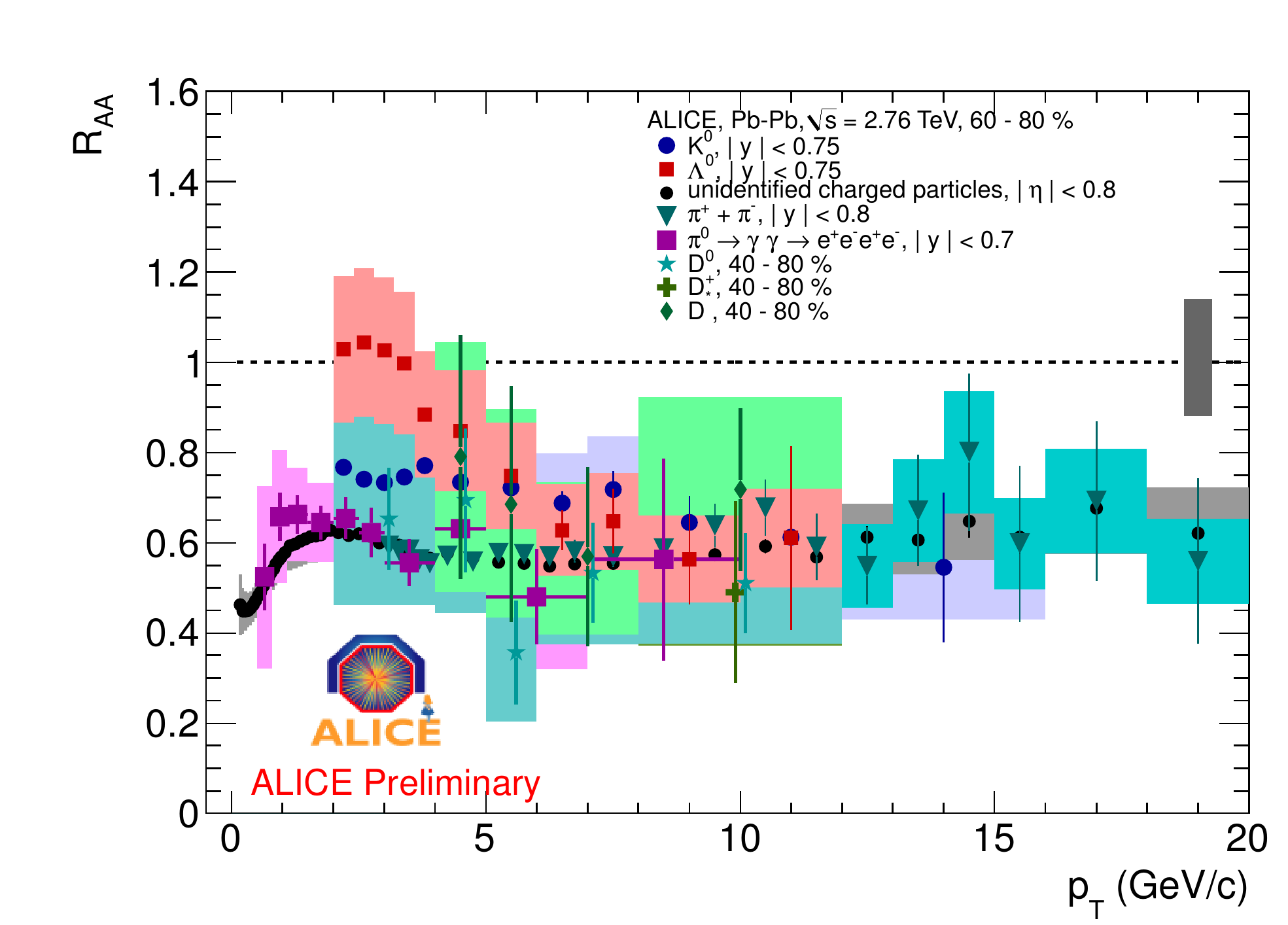}
  \caption{Nuclear modification factor $R_{AA}$ of charged particles,
    $K^0$, $\Lambda$, $\pi^\pm$, $D^+$, $D^0$, $D^{*+}$ in central
      (left) and peripheral (right) collisions.}
  \label{fig:PbPb-RAA_all}
\end{figure}

A nuclear modification factor $R_{AA}$ of charged pion production in
mid-rapidity (Fig.\ref{fig:PbPb-RAA_pions}) has lower values in the
range of moderate transverse momenta ($3<\pT<7-10$~GeV/$c$) than that
of unidentified charged particles, but at higher $\pT$ it coincides with
all charged particles.
To the contrary to charged pions, strange hadrons ($K^0_S$, $\Lambda$)
are less  suppressed in  the most central  collisions compared  to all
charged  particles (Fig.\ref{fig:PbPb-RAA_all}).  This is
explained by the  fact that strange quark production  is enhanced in a
hot  nuclear  medium,   and  this  strangeness  enhancement  partially
compensates energy loss of strange quarks, such that the overall
$R_{AA}$ value becomes larger than for pions. Lambda hyperons have no
suppression at $\pT<3-4$~GeV/$c$, which is interpreted by an
additional baryon enhancement in central heavy-ion collisions.

ALICE has reported also the first measurements of $D$ meson
suppression \cite{bib:PbPb-Dmesons} in Pb-Pb collisions in two
centrality classes, $0-20\%$ and $40-80\%$, shown in
Fig.\ref{fig:PbPb-RAA_all}. It was shown that the $R_{AA}$ values for
$D^0$, $D^+$ and $D^{*+}$ are consistent with each other within the
statistical and systematical uncertainties. Although the statistics of
the ALICE run 2010 is marginal for $D$ meson measurement, the obtained
result shows a hint that the $D$ mesons are less suppressed than
charged pions.

Nuclear modification factor $R_{AA}$ is $J/\psi$ production in Pb-Pb
collisions was measured by the ALICE in two kinematic regions: in the
forward rapidity with the muon spectrometer and in mid-rapidity
deploying the central tracking system \cite{bib:PbPb-Jpsi}. The result
of $R_{AA}$ measurement in forward rapidity shows almost no dependence
on collision centrality with the average value $R_{AA} = 0.545 \pm
0.032~\mbox{(stat.)} \pm 0.084 \mbox{(syst.)}$ which is significantly
different from the RHIC results.

%-----------------------------------------------------------------------------
\section{Conclusion}

The ALICE collaboration is performing QCD studies via hadron
production measurements in proton-proton collisions. Obtained results
in pp collisions at $\sqrt{s}=0.9, 2.76$ and 7~TeV show
statistically significant deviations from models which well described
lower-energy results. Therefore new experimental results from pp
collision allow to tune various phenomenological models and pQCD
calculations.

Comprehensive studies of heavy-ion collisions, performed by the ALICE
experiment show that the properties of strongly interacting nuclear
matter produced at the LHC energy, qualitatively similar to those
observed at RHIC and reveal smooth evolution with collision
energy. The matter produced at LHC has about 3 times larger energy
density, twice larger volume of homogeneity and about 20\% larger
lifetime. Like at RHIC, the matter at LHC reveals the properties on an
almost perfect liquid. Particle suppression appeared to be stronger at
LHC than at RHIC which is also an evidence of denser medium produced
at LHC.

This work was parially supported by the RFBR grant 10-02-91052.

\newpage
%-----------------------------------------------------------------------------

%
%
\end{document}